\documentstyle{mn}
\input{epsf}

\newif\ifAMStwofonts
\AMStwofontstrue

\def\be{\begin{equation}}
\def\ee{\end{equation}}
\def\go{\mathrel{\raise.3ex\hbox{$>$}\mkern-14mu
             \lower0.6ex\hbox{$\sim$}}}
\def\lo{\mathrel{\raise.3ex\hbox{$<$}\mkern-14mu
             \lower0.6ex\hbox{$\sim$}}}
\def\br{{\bf r}}
\def\vxi{{\vec\xi}}

\title{Secular Instability of g-Modes in Rotating Neutron Stars}
\author[Dong Lai]
       {Dong Lai\\
        Centre for Radiophysics and Space Research,
	Department of Astronomy, 
	Cornell University, Ithaca, NY 14853, USA\\
	{\rm E-mail: dong@spacenet.tn.cornell.edu}}
\date{Accepted 1999,4; 
      Received 1998 xxx;
      in original form 1998 xxx}

\pagerange{\pageref{firstpage}--\pageref{lastpage}}
\pubyear{1999}

\begin{document}

\maketitle

\label{firstpage}

\begin{abstract}
Gravitational radiation tends to drive gravity modes in rotating 
neutron stars unstable. For an inviscid star, 
the instability sets in when the rotation frequency
is about $0.7$ times the corresponding mode frequency of the nonrotating
star. Neutron stars with spin frequencies $\go 100$ Hz are susceptible 
to this instability, with growth time of order years.  
However, it is likely that viscous dissipation suppresses the 
instability except for a narrow range of temperatures around $10^9$~K. 
We also show that the viscosity driven instability of g-modes is absent.
\end{abstract}


\section{Introduction}

It is well known that globally nonaxisymmetric instabilities can develop
in rapidly rotating neutron stars. Of particular interest is the secular 
instability driven by gravitational radiation,
as first discovered by Chandrasekhar (1970) for the 
bar-mode ($m=2$) of an incompressible Maclaurin spheroid, and later shown by 
Friedman \& Schutz (1978a,b) to be a generic feature of rotating stars. 
The Chandrasekhar-Friedman-Schutz instability (hereafter CFS 
instability) occurs whenever a backward-going mode 
(with respect to the rotation) in the corotating frame of the star
is ``dragged'' by rotation to become forward-going in the inertial frame.
A CFS unstable mode has negative energy (in the inertial frame), 
and gravitational radiation makes the energy even more negative.

The CFS instability has been extensively studied for f-modes
since it is widely believed that the maximum angular velocities of
neutron stars are determined by instabilities of these modes
(e.g., Ipser \& Lindblom 1990, 1991; Cutler \& Lindblom 1992;
Lindblom 1995; Yoshida \& Eriguchi 1995; Stergioulas \& Friedman 1998).
The f-mode instabilities occur at rotation frequencies comparable to 
the maximum ``break-up'' frequency ($\sim 1000$~Hertz). It was found that 
viscous dissipation in the neutron star stablizes almost all f-modes
except in certain limited temperature regimes. 
Recently Andersson (1998) pointed out that r-modes 
(previously studied by Papaloizou \& Pringle 1978, Provost et al.~1981,
Saio 1982, etc.) are also destablized by gravitational radiation 
(see also Friedman \& Morsink 1998). Moreover, the couplings between 
gravitational radiation and r-modes (mainly through current multipole 
moments) are so strong that viscous forces present in hot young neutron stars
are not sufficient to suppress the instability (Lindblom et al.~1998;
Andersson et al.~1998a). It is therefore likely that gravitational
radiation can carry away most of the angular momentum of a nascent 
neutron star through the unstable r-modes, and the gravitational waves 
generated in this process could potentially be detectable by 
LIGO (Owen et al.~1998). The implications of the r-mode instability for 
accreting neutron stars have also been discussed (Andersson et al.~1998b;
Bildsten 1998; Levin 1998).

A neutron star also possesses p-modes and g-modes, whose stability properties 
have not been studied. The p-modes are similar to the f-modes, except 
that, as compared to the f-modes, they have higher frequencies, 
weaker couplings to the gravitational radiation and stronger 
viscous dampings. Thus the p-mode instabilities, if exist, 
are certainly less important than the f-mode instabilities.
The situation with g-modes is not clear, however. Gravity modes in neutron
stars arise from composition (proton to neutron ratio)
gradient in the stellar core (Reisenegger \& Goldreich 1992; Lai 1994),
density discontinuities in the crust (Finn 1987; Strohmayer 1993) as well as
thermal buoyancy associated with finite temperatures, either due to
internal heat (McDermott, Van Horn \& Hansen 1988) or due to
accretion (McDermott \& Taam 1987; Bildsten \& Cutler 1995; 
Strohmayer \& Lee 1996; Bildsten \& Cumming 1998). 
For a nonrotating neutron star, 
the typical frequencies of low-order g-modes are around $100$~Hz and  
are smaller for higher-order modes. Such low frequencies imply
that, contrary to f-modes, g-modes may be destablized by gravitational 
wave when the star rotates at a rate much smaller than the break-up spin
frequency. Moreover, since g-modes possess small, 
but nonzero mass quadrupole moments, 
it is not clear {\it a priori} whether the driving due to gravitational
radiation is stronger or weaker for g-modes than that for r-modes 
(which have zero mass quadrupole moment). The purpose of this paper is to 
address these issues related to the secular instability of
g-modes. 

Our paper is organized as follows. We start in Section 2 by summarizing
the essential properties of g-modes in nonrotating neutron stars and defining
notations. 
In Section 3 we calculate the effect of rotation on the mode frequencies and
determine the onset of CFS instabilities of g-modes for inviscid stars.  
In Section 4 we evaluate the driving/damping rates of the g-modes due to
gravitational radiation and viscosities. Section 5 considers 
the possibility of viscosity-driven g-mode instability. 
In Section 6 we compare the instabilities of g-modes and 
other modes (f and r-modes) and discuss some possible astrophysical 
applications of our results.

\section{Preparations: Core g-modes of Neutron Stars}

In this paper, we shall focus on the core g-modes, with buoyancy
provided by the gradient of proton to neutron ratio in the neutron star
interior (Reisenegger \& Goldreich 1992). These modes exist
even in zero-temperature stars.  
The other types of g-modes (due to crustal density discontinuities or thermal
entropy gradient) either have weaker restoring forces or are 
confined to the outer region of the star and therefore are 
less susceptible to the CFS instabilities as compared to the core
g-modes (see below).  

We construct equilibrium neutron star models by solving Newtonian
hydrostatic equations. Centrifugal distortion to the stellar structure
is neglected as we shall only consider stars with rotation rates far below
the break-up value. The equations of state (EOS) used for
our neutron star models and g-mode calculations have been described
in Lai (1994). The nuclear symmetry energy plays an important role
in determining the proton fraction $x=n_p/n$ (where $n$ is the baryon number 
density and $n_p$ in the proton number density) and hence the
frequencies of g-modes. 
Stable stratification of the liquid core (thus the existence of 
core g-modes) arises from the $x$-gradient, and from the fact the
$\beta$-equilibration time is much longer than the dynamical times
of the modes. The Brunt-V\"ais\"al\"a (angular)
frequency $N$ is given by
\be
N^2=-{g\over\rho}\left({\partial\rho\over\partial x}\right)_P\left({dx\over
dr}\right),
\label{n2}\ee
where $g>0$ is the local gravitational accelaration.
As an estimate of the mode frequency, we neglect the nucleon interaction, 
and find $x\simeq 5.6\times 10^{-3}(\rho/\rho_{\rm nuc})$, where 
$\rho_{\rm nuc}=2.8\times
10^{14}$~g~cm$^{-3}$ is the nuclear density. The Brunt-V\"ais\"al\"a frequency
is then
\be
N=g\left({3\rho\over 10P}x\right)^{1/2}\sim (G\rho x)^{1/2}
\sim 10^3~{\rm s}^{-1}.
\ee
Thus the core g-modes typically have frequencies less than 
$100-200$~Hz. 

\begin{figure}
\vskip -1.2 truecm
\epsfxsize=10.2cm
\epsfysize=10.2cm
\hbox{\hskip -1.0 truecm \epsfbox{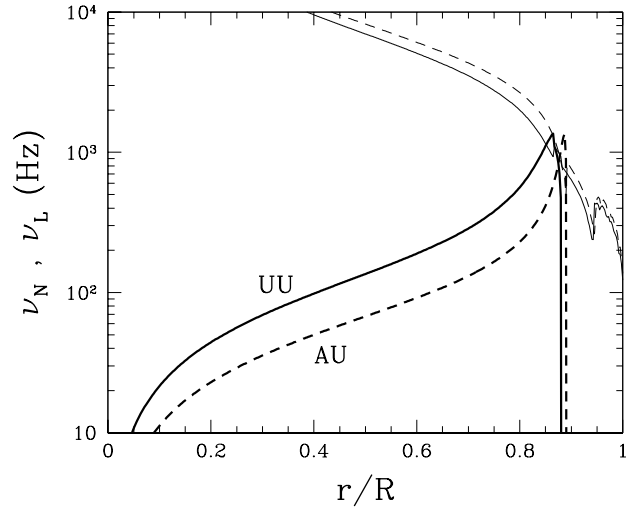}}
\vskip -1.9 truecm
\caption{Profiles of the Brunt-V\"ais\"al\"a frequency $\nu_N$ (heavy curves)
and the Lamb frequency $\nu_L$ (light curves) for two neutron star 
models: UU (solid curves) and AU (dashed curves). We have
set $\nu_N=0$ outside the core ($\rho\lo 10^{14}$~g~cm$^{-3}$). 
The dips in the $\nu_L$ curves at $r/R\simeq 0.94$ 
correspond to neutron drip at $\rho \simeq 4\times 10^{11}$~g~cm$^{-3}$.} 
\end{figure}

In the following sections, we shall consider two representative models 
based on a parametrized EOS derived from Wiringa, Fiks \& Fabrocini (1988), 
and labeled as UU and AU in Lai (1994). Figure 1 depicts the profiles of 
the Brunt-V\"ais\"al\"a frequency, $\nu_N=N/(2\pi)$, as a function of
radius inside the star for the two models. We see that $\nu_N=0$ 
at the center (where $g=0$) and increases outward because of 
decreasing sound speed $c_s$ and increasing $g$ (despite that $x$ decreases
outward). We have set $\nu_N=0$ outside the core ($\rho\lo
10^{14}$~g~cm$^{-3}$)\footnote{The transition from the core (where 
only free nucleons and electrons exist) to the inner crust (where heavy
nuclei co-exist with free neutrons) is somewhat ambiguous. The inner crust is
described by the equation of state of Baym, Bethe \& Pethick
(1971) while the core is described by that of Wiringa et al.~(1988).  
We have found the g-mode properties to be insensitive to the
actual transition radius within the allowed range.}, 
and thus eliminated the crustal modes (see below). 
For completeness, Figure 1 also shows the Lamb frequency (or
the acoustic cut-off frequency), $\nu_L=\sqrt{l(l+1)}\,c_s/(2\pi r)$, 
with $l=2$. Gravity wave (with frequency $\nu$)
can propagate only in the region satisfying 
$\nu<\nu_L$ and $\nu<\nu_N$. For a typical $\nu\sim 100$~Hz, the wave
propagation region lies in the range $0.5\lo r/R\lo 0.9$.
For convenience, the basic properties of
the modes for nonrotating stars are summarized in Table 1. 
\begin{table*}
 \centering
 \begin{minipage}{140mm}
  \caption{Core g-modes of nonrotating neutron stars}
  \begin{tabular}{ccccccc}
   Model & $M~(M_\odot)$ & $R$~(km) & $\rho_c$~(g~cm$^{-3}$) & mode
\footnote{The subscript $n$ in g$_n$ specifies the radial order of the mode.
The angular order is $l=2$.}
     & $\nu$~(Hz) & $|\delta D_{22}|~$\footnote{Note that the values 
of $|\delta D_{22}|$ quoted here
differ from those in Lai (1994) by up to $80\%$. This is because we adopt 
Cowling approximation in this paper while in Lai (1994) the full equations 
are solved; see discussion at the end of \S 3.}
\\[9pt]
 UU & 1.4 & 13.48 & $6.95\times 10^{14}$ & g$_1$ &148 & $6.2\times 10^{-4}$\\
    &  & & & g$_2$ & 97 & $1.8\times 10^{-5}$ \\
 AU & 1.4 & 12.37 & $8.18\times 10^{14}$ & g$_1$ & 72 & $1.1\times 10^{-4}$\\
    &  & & & g$_2$ &52 & $9.9\times 10^{-6}$ \\
\end{tabular}
\end{minipage}
\end{table*}

The crustal discontinuity modes have angular frequencies approximately 
given by $\omega\sim (g/R)^{1/2}(\Delta\rho/\rho)^{1/2}$. With typical density
discontinuity $\Delta\rho/\rho\sim 1\%$, this gives a frequency
of order $200$~Hz (Finn 1987).
Although these modes can have frequencies greater than the core g-modes,
the crustal modes are confined in the outer region of the star
and therefore have very small mass quadrupole moments\footnote
{This is indicated by the fact that the mode damping time due to
gravitational radiation is rather long, exceeding $10^4-10^5$~years.}.
The thermally driven g-modes (McDermott et al.~1988)
have frequencies less than $50$~Hz for $T\lo 10^{10}$~K, 
indicating smaller restoring forces as compared to
the core g-modes. We will not consider the crustal modes and thermal modes
in the rest of this paper.

\section{Onset of Instability: Inviscid Stars}

For neutron stars consisting of ideal fluid (neglecting
viscosity), it is possible to determine the onset of CFS instability
of g-modes without calculating the growth/damping rates
due to dissipative processes.
Consider a mode with azimuthal angular dependence $e^{im\phi}$
and time dependence $e^{i\omega_r t}$, where $\omega_r$ ($>0$) is the 
mode angular frequency in the rotating 
frame. The rotation angular frequency is $\Omega_s$ ($>0$). A mode with 
$m>0$ propagates in the direction opposite to the rotation,
with pattern speed $(-\omega_r/m)$ (in the rotating frame). In the 
inertial frame, the mode frequency is 
\be
\omega_i=\omega_r-m\Omega_s.
\ee
When $\omega_i<0$, the mode attains a prograde pattern speed,
$-(\omega_i/m)$, in the inertial frame, and the CFS instability
sets in (Friedman \& Schutz 1978a,b).  

Since we shall consider $\Omega_s$ comparable to $\omega_r$, a
perturbative treatment of the rotational effect is inadequate
(see the last paragraph of this section).
We shall adopt the so-called ``traditional approximation''
(Chapman \& Lindzen 1970; Unno et al.~1989;
Bildsten et al.~1996), where the Coriolis forces associated with the
horizontal component of the spin angular velocity are neglected.
This allows the mode eigenfunction to be separated into angular and radial 
components.
Note that the traditional approximation is strictly valid only 
for high-order g-modes for which $\omega_r\ll N$ 
and $\xi_h\gg \xi_r$ (where $\xi_h\sim \xi_\theta\sim \xi_\phi$
is the horizontal Lagrangian displacement and $\xi_r$ is the
radial displacement), and for
$\Omega_s\ll N$ (so that the radial component of the Coriolis force can be
neglected compared to the buoyancy force). While these conditions 
are not valid everywhere inside the star (e.g., near the center $\xi_r$
is of the same order as $\xi_h$, and $\omega_r,\Omega_s>N$; also, near the
surface, $\omega_r,\Omega_s>N$), in the region 
($0.6\lo r/R\lo 0.9$) where the mode propagates,
they are reasonably satisfied (see below).

Neglecting the perturbation in gravitational potential (Cowling
approximation), the radial Lagrangian displacement and
Eulerian pressure perturbation can be written as
\be
\xi_r(\br)=\xi_r(r)H_{jm}(\theta)e^{im\phi},~~~
\delta P(\br)=\delta P(r)H_{jm}(\theta)e^{im\phi},
\label{eigen}\ee
where $H_{jm}(\theta)$ is the Hough function 
($jm$ are the indices analogous to $lm$ in the spherical harmonic
$Y_{lm}$), satisfying the
Laplace tidal equation (Chapman \& Lindzen 1970; Bildsten et al.~1996):
\begin{eqnarray}
&&{\partial\over\partial\mu}\left({1-\mu^2\over 1-q^2\mu^2}{\partial H_{jm}
\over
\partial\mu}\right)-{m^2H_{jm}\over (1-\mu^2)(1-q^2\mu^2)}\nonumber\\
&&~~~~~~~~~~~~~~~~~~~~~~~-{qm(1+q^2\mu^2)H_{jm}\over 
(1-q^2\mu^2)^2}=-\lambda H_{jm},
\label{laplace}\end{eqnarray}
with $\mu=\cos\theta$ and $q=2\Omega_s/\omega_r$. 
The eigenvalue, $\lambda$, depends
on $m$ and $q$. For $q\rightarrow 0$ (the nonrotating case), 
the function $H_{jm}(\theta)e^{im\phi}$ becomes $Y_{lm}(\theta,\phi)$ while
$\lambda$ degenerates into $l(l+1)$.
The other eigenfunctions of the mode are related to $\delta P$ and $\xi_r$ via
\begin{eqnarray}
\delta\rho(\br) &=&\delta\rho(r)H_{jm} e^{im\phi}
=\left({\delta P\over c_s^2}+{\rho N^2\over g}\xi_r\right)
H_{jm} e^{im\phi},\\
\xi_\theta(\br) &=& {\xi_\perp(r)\over 1-q^2\mu^2}
\left({\partial H_{jm}\over\partial\theta}
+{mq\mu\over\sin\theta}H_{jm}\right)e^{im\phi},\\
\xi_\phi(\br) &=&{i\xi_\perp(r)\over 1-q^2\mu^2}
\left(q\mu{\partial H_{jm}\over\partial\theta}
+{m\over\sin\theta}H_{jm}\right)e^{im\phi},
\end{eqnarray}
where $c_s$ is the sound speed and
$\xi_\perp(r)=\delta P(r)/(\rho r\omega_r^2)$.
Separating out the angular dependence, the fluid continuity equation
and Euler equation (in the rotating frame)
reduce to a set of coupled radial equations:
\begin{eqnarray}
{d\over dr}\delta P &=&-{g\over c_s^2}\delta
P+\rho(\omega_r^2-N^2)\xi_r,\label{eqdelp}\\
{d\over dr}(r^2\xi_r) &=& {g\over c_s^2}(r^2\xi_r)
-{r^2\over c_s^2\rho}\delta P+{\lambda\over\omega_r^2\rho}\delta P.
\label{eqxir}
\end{eqnarray}

The properties of the Hough function have been extensively studied
(Longuet-Higgins 1967). We have adopted
the numerical approach of Bildsten et al.~(1996)
in our calculation. For given $q$ and $m$, the eigenvalue
$\lambda$ is obtained by solving equation (\ref{laplace}). To 
obtain the eigen-frequency $\omega_r$, we solve 
the radial equations (\ref{eqdelp})-(\ref{eqxir}) subjected
to the following boundary conditions: (i) At the surface $r=R$,
we have the standard requirement $\Delta P=\delta P-\rho g\xi_r=0$;
(ii) At a small $r$ close the the center,
regularity requires $\delta P/\rho=Y_0 r^{\beta}$,
$\xi_r=({\beta}/\omega_r^2)Y_0\,r^{\beta-1}$ ($Y_0$ is a constant),
where ${\beta}$ is determined from $\beta (1+\beta)=\lambda$, or
$\beta=\left(\sqrt{1+4\lambda}-1\right)/2$.
Once $\omega_r$ is found, the actual rotation rate $\Omega_s$ 
is recovered from $\Omega_s=q\omega_r/2$. 

Figure 2 shows the frequencies (in the inertial frame) of the 
$j=m=2$ g$_1$ modes for model AU and UU as functions of the spin
frequencies (These modes have $l=2$ in the
$\Omega_s=0$ limit). Note that as $\Omega_s$ increases, $\omega_r$ increases 
while $\omega_i$ decreases. 
\begin{figure}
\vskip -1.2 truecm
\epsfxsize=10.2cm
\epsfysize=10.2cm
\hbox{\hskip -1.0 truecm \epsfbox{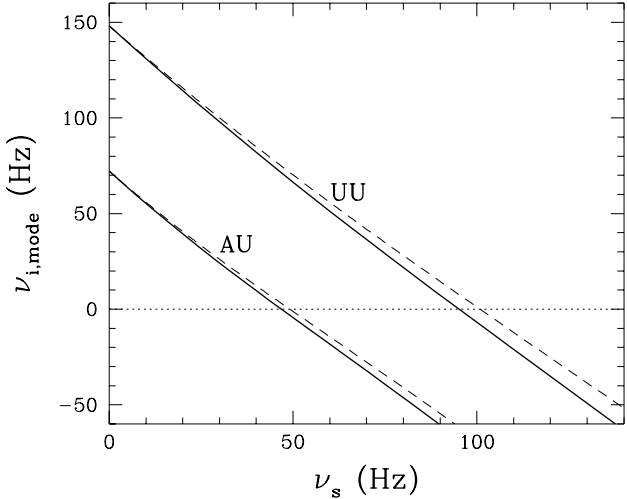}}
\vskip -1.9 truecm
\caption{The frequencies of the $j=m=2$ g$_1$ modes in the inertial frame
for model AU and UU as functions of the spin frequency. The onset of
instability occurs at $\nu_{i,{\rm mode}}=0$. The solid lines are numerical
results obtained by solving the radial equations; The dashed lines
correspond to the WKB limits.}
\end{figure}

An approximate solution the the radial equation can be obtained by
the WKB analysis. Substituting $r^2\xi_r\propto e^{ikr}$ and 
$\delta P/\rho\propto e^{ikr}$ into the radial equations, we obtain
the dispersion relation for g-modes:
\be 
k^2\simeq  {\lambda\over\omega_r^2 r^2}(N^2-\omega_r^2)
\ee
This indicates that $\omega_r\propto\sqrt{\lambda}$. In fact, with the
condition $\int_{r_1}^{r_2}k\,dr=(n+C)\pi$ ($n$ is the 
number of radial node, $C$ is a constant of order unity, 
$r_1$ and $r_2$ are the inner and outer turning points where 
$\omega_r=N$), we can write
\be
\nu_r\simeq {234\,{\rm Hz}\over n+C}\left({\lambda\over 6}\right)^{1/2}
\left({\int\!\nu_N\,d\ln r\over 300\,{\rm Hz}}\right).
\ee
Thus we have,
for the $j=2$ modes,
$\omega_r/\omega^{(0)}=\sqrt{\lambda/6}$ and
$\Omega_s/\omega^{(0)}=q\sqrt{\lambda/24}$,
where $\omega^{(0)}$ is the corresponding mode frequency at zero
rotation. The WKB results are also
plotted in Fig.~2, showing remarkable agreement with the actual numerical
result. Clearly, $\omega_i=0$ occurs at $q=1$,
which corresponds to $\lambda=11.13$ from the solution of the
$m=2$ Laplace equation (eq.~[\ref{laplace}]). Thus, the onset of instability 
occurs when the ratio of spin frequency $\Omega_s$ and the nonrotating mode
frequency $\omega^{(0)}$ satisfies
\be
{\Omega_s\over\omega^{(0)}}=\left({11.13\over 24}\right)^{1/2}=0.68
\label{critical}\ee
The actual numerical calculations give slightly smaller critical 
spin frequency (see Fig.~2). 
For $\bar\Omega_s\equiv \Omega_s/\omega^{(0)}<1$, in the WKB limit, 
the $j=m=2$ mode frequency can be fitted to the following
analytic expression to within $1\%$:
\footnote{
A series solution (in $q$) of eq.~(\ref{laplace}) yields
$\lambda=6+2q+{10q^2/7}+\cdots$ for $j=m=2$
(Bildsten \& Ushomirsky 1996, private communication). 
In eq.~(\ref{omegai}), the linear and quadratic terms are derived from 
this analytic expansion of $\lambda$ for small $q$ 
and the last term is based on our numerical fitting.}
\be
{\omega_i\over\omega^{(0)}}=1-{5\over 3}{\bar\Omega_s}
+{13\over 42}{\bar\Omega_s}^2-0.064\,{\bar\Omega_s}^{4.6}.
\label{omegai}\ee

\begin{figure}
\vskip -1.2 truecm
\epsfxsize=10.2cm
\epsfysize=10.2cm
\hbox{\hskip -1.0 truecm \epsfbox{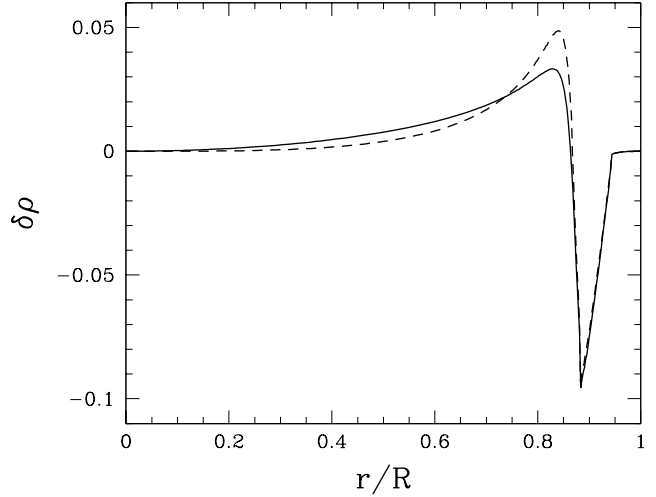}}
\vskip -1.9 truecm
\caption{Density perturbation $\delta\rho$ (in arbitrary unit) 
as a function of radius
for the $j=m=2$ g$_1$ mode of model UU. The solid line is 
for $\nu_s=0$, and the dashed line is for $\nu_s\simeq 140$~Hz 
(corresponding to $q=1.3$).} 
\end{figure}

Figure 3 gives two examples of the radial eigenfunctions.
We see that the mode is concentrated in the outer
region of the stellar core ($0.6\lo r/R\lo 0.9$), and rotation tends
to ``squeeze'' the mode into that region. This is because rotation increases
$\omega_r$ and narrows the region in which the wave can propagate 
($\omega_r<N$). The dispacements $\xi_r$ and $\xi_\perp$ can be
quite significant near the surface, but the associated energy density
(proportional to $\rho|\vec\xi|^2$) is significant only in the
propagation zone. 

We can similarly calculate the frequencies of higher-order g-modes.
The results are even closer to the WKB limit (as expected). 
These modes tend to be unstable at lower spin frequencies.
Similarly, the higher-$m$ modes are also destablized by gravitational
waves. We shall not be concerned with these higher-order
modes since they have smaller growth rates due to gravitational
radiation and higher damping rates due to viscosities as
compared to the $m=2$,~g$_1$ mode (see Section 4).  

Finally, we discuss the validity of the two approximations adopted in our 
calculations. As far as the mode frequency is concerned, the 
Cowling approximation is an excellent approximation for g-modes:
For nonrotating stars, our mode frequencies agree with those
of Lai (1994) (where the full equations are solved) to within $2\%$. 
The Cowling approximation does introduce an error (up to a factor of
$80\%$) to the quadrupole moment of the mode (see \S 4). However, 
the uncertainty in the nuclear equation of state (particularly
the nuclear symmetry energy which determines the g-mode
properties) is presumably larger. 
As noted before, the conditions $\omega_r\ll N$ and $\Omega_s\ll N$,
needed for the validity of the traditional approximation, 
are satisfied in the propagation zone of the mode, although they 
are clearly violated near the stellar center and surface. 
Our numerical results indicate that for the g$_1$ mode, the 
inequality $|\xi_\perp|>|\xi_r|$ is achieved only for $r/R\go 0.6-0.7$ 
(depending on the rotation rate),
and $|\xi_\perp|\gg |\xi_r|$ is achieved only near the stellar surface
[indeed $\xi_\perp/\xi_r=GM/(R^3\omega_r^2)\gg 1$ at $r=R$]. 
Thus the traditional approximation is not rigorously justified for the
g$_1$ mode, although it becomes increasingly valid for higher-order modes.  
Clearly, without an exact, full calculation, which is currently beyond reach, 
it is difficult to know definitively how good the traditional approximation is
when applying to the low-order modes. 
One way to assess the validity of our results is 
to consider the $\omega^{(0)}\gg\Omega_s$
limit using the linear theory, which gives, for the $l=m=2$ mode:  
\be
\omega_i=\omega^{(0)}-2(1-C_{nl})\Omega_s,
\label{linear}\ee
where
\be
C_{n2}={\int_0^R\!\rho\,r^2\,(2\xi_r\xi_\perp+\xi_\perp^2)\,dr\over
\int_0^R\!\rho\,r^2\,[\xi_r^2+6\xi_\perp^2]\,dr}
\ee
(e.g., Unno et al.~1989). If we assume $|\xi_\perp|\gg |\xi_r|$,
we find $C_{nl}=1/6$ and eq.~(\ref{linear}) reduces to the first two
terms of eq.~(\ref{omegai}). For model UU, our numerical calculation yields
$C_{nl}=0.104,~0.132,~0.148,~0.155$ for 
g$_1$,g$_2$,g$_3$,g$_4$ respectively, corresponding to
$2(1-C_{nl})=1.79,~1.74,~1.70,~1.69$. These values are close
to the corresponding value ($5/3=1.67$)
obtained from the traditional approximation.  
Since our calculation indicates that the instability occurs 
at $\Omega_s<\omega^{(0)}$ (see eq.~[\ref{critical}]), we are confident
that our qualitative conclusion regarding the secular instability of g-modes
is firm; the critical ratio $\Omega_s/\omega^{(0)}$ for instability 
obtained in this paper (eq.~[\ref{critical}]) is likely to be
accurate to within $\sim 20\%$.

\section{Driving and Damping of the Modes}

The condition that $\omega_i$ changes sign is only a necessary condition
for the CFS instability. For the instability to be of interest,
it must grow sufficiently fast and that the growth rate must overcome
the possible damping rates due to viscosities. In this section we
determine the mode growth rate and damping rate. 

The timescale $\tau$ associated with any dissipative process is given by
\be
{1\over\tau}=-{1\over 2E}{dE\over dt},
\ee
where $E$ is the canonical mode energy in the rotating frame
(Friedman \& Schutz 1978a):
\begin{eqnarray}
E&=&{1\over 2}\int\!\!d^3\!x\!\biggl[\rho\,\omega_r^2\,\vxi^\ast\cdot\vxi
+\left({\delta P\over\rho}-\delta\Phi\right)\delta\rho^\ast\nonumber\\
&&+(\nabla\cdot\vxi)\,\vxi^\ast\cdot (\nabla p-c_s^2\nabla\rho)\biggr].
\label{energy}\end{eqnarray}
We normalize the mode amplitude according to
\be
\int\!d^3\!x\,\rho\,\vxi^\ast\cdot\vxi
=\int_0^Rdr\,r^2\rho\left(\xi_r^2+\Lambda\xi_\perp^2\right)=1,
\ee
where $\Lambda$ is evaluated using
\begin{eqnarray}
\Lambda&=&2\pi\int_{-1}^1{d\mu\over (1-q^2\mu^2)^2}
\biggl[(1+q^2\mu^2)(1-\mu^2)
\left({\partial H_{jm}\over\partial\mu}\right)^2\nonumber\\
&&+{m^2(1+q^2\mu^2)\over 1-\mu^2}H_{jm}^2-4mq\mu H_{jm}
\left({\partial H_{jm}\over\partial\mu}\right)\biggr],
\end{eqnarray}
and the Hough function is normlized via $2\pi\int\!d\mu\,H_{jm}^2=1$.
The g-mode energy is dominated by the first term in eq.~(\ref{energy}),
which gives
\be
E\simeq {1\over 2}\omega_r^2 (1+{\cal E}),
\ee
with
\be
{\cal E}={1\over\omega_r^2}\int\!dr\,r^2{\delta P(r)\over\rho}\delta\rho(r).
\ee

The dissipation due to gravitational waves is calculated from the 
multipole radiation formula derived by Lindblom et al.~(1998)
using the formalism of Thorne (1980):
\be
\left({dE\over dt}\right)_{\rm gr}=-\omega_r\sum_{l\ge 2}
N_l\omega_i^{2l+1}(|\delta D_{lm}|^2+|\delta J_{lm}|^2).
\ee
where $N_l$ is the coupling constant:
\be
N_l={4\pi G\over c^{2l+1}}{(l+1)(l+2)\over l(l-1)[(2l+1)!!]^2},
\ee
and the mass and current multipole moments of the perturbed fluid are
given by:
\be
\delta D_{lm}=\int\!\!d^3\!x\,\delta\rho\,r^lY_{lm}^\ast,
\ee
\be 
\delta J_{lm}={2\over c}\sqrt{l\over l+1}\int\!\!d^3\!x\,
r^l(\rho i\omega_r\vxi+\delta\rho {\bf v})\cdot
{\vec Y}_{lm}^{B\ast},
\ee
and ${\vec Y}_{lm}^B$ is the magentic type vector spherical harmonic
defined by Thorne (1980):
\be
{\vec Y}_{lm}^B={1\over \sqrt{l(l+1)}}r\nabla\times (r\nabla Y_{lm}).
\ee

For the $j=m=2$ mode, the dominant gravitational radiation comes from the mass
quadrupole $\delta D_{22}$. We find
\be
{1\over\tau_{\rm gr}}\simeq {25\over 1+{\cal E}}
\hat\omega_i^5\hat\omega_r^{-1}
\left({\delta D_{22}\over 10^{-4}}\right)^2M_{1.4}^3R_1^{-4}~{\rm yr}^{-1},
\ee
where $\hat\omega=\omega/(GM/R^3)^{1/2}$, $\delta D_{22}$ is dimensionless
(in units where $G=M=R=1$), $M_{1.4}=M/(1.4\,M_\odot)$ and
$R_1=R/(10\,{\rm km})$.
Clearly, $\tau_{\rm gr}$ changes from positive (damping) for
$\omega_i>0$ to negative (driving) for $\omega_i<0$ (where the mode
is unstable for inviscid stars).

Figure 4 shows the damping/growth rate $\tau_{\rm gr}^{-1}$
as a function of the stellar spin frequency $\nu_s$
for the g$_1$ mode of the two neutron star models. The change
of $\tau_{\rm gr}^{-1}$ with varying $\nu_s$ mainly comes from 
the changes in $\omega_i$ and $\omega_r$, with the change in $\delta D_{22}$
relatively minor. We see that when $\nu_s$ exceeds the critical spin rate for
instability, the mode growth rate can be significantly larger than
the damping rate for the nonrotating models (This is particularly 
evident for model AU). 

\begin{figure}
\vskip -1.2 truecm
\epsfxsize=10.2cm
\epsfysize=10.2cm
\hbox{\hskip -1.0 truecm \epsfbox{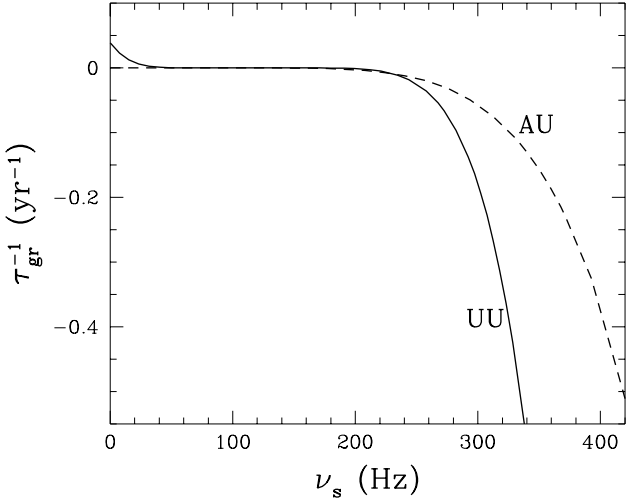}}
\vskip -1.9 truecm
\caption{The gravitational radiation damping rate $\tau_{\rm gr}^{-1}$
as a function of spin frequency for the two g-modes depicted in Fig.~2.
Positive $\tau_{\rm gr}^{-1}$ (for small $\nu_s$) 
implies damping, while negative $\tau_{\rm gr}^{-1}$ implies driving of the 
mode by gravitational radiation. At $\nu_s=0$,
we have $\tau_{\rm gr}^{-1}\simeq 0.04$~yr$^{-1}$ for model UU and
$\tau_{\rm gr}^{-1}\simeq 6\times 10^{-5}$~yr$^{-1}$ for model AU.
}
\end{figure}

The higher order mass multipole radiations (from $\delta D_{l2}$, with
$l>2$) are less important. One can show that the radiation from
$\delta J_{22}$ identically vanishes. The leading order
current multipole radiation comes from $\delta J_{32}$, but 
it is much smaller than the mass quadrupole radiation. 

The growth/damping timescale of g-modes due to 
gravitational radiation is typically much longer than that of
f-modes (for which $\tau_{\rm gr}$ is of order seconds). This
comes about for two reasons: (i) G-modes have lower frequencies;
(ii) The multipole moment $\delta D_{lm}$ for g-modes is much smaller
than that for the f-modes 
(for which $\delta D_{lm}$ is of order unity).
\footnote{The small $\delta D_{lm}$ for g-modes can be understood from 
the sum rule (Reisenegger 1994) that must be satisfied for the multipole 
moments of all modes.}

The dissipation of mode energy due to viscosities can be calculated from
(Ipser \& Lindblom 1991)
\be
\left({dE\over dt}\right)_{\rm visc}=-\int\!\!d^3\!x\,\left(2\eta
\,\delta\sigma^{ab}\delta\sigma_{ab}^\ast+\zeta|\delta\sigma|^2\right),
\ee
where $\delta\sigma_{ab}$ and $\delta\sigma$ are the shear and expansion 
of the perturbation, respectively:
\begin{eqnarray}
\delta\sigma_{ab}&=&{i\omega_r\over 2}\left(\nabla_a\xi_b+\nabla_b\xi_a-
{2\over 3}\delta_{ab}\nabla_c\xi^c\right),\\
\delta\sigma &=& i\omega_r\nabla_c\xi^c.
\end{eqnarray}
The shear viscosity resulting from neutron-neutron scattering in the normal
liquid core is given by (Flowers \& Itoh 1979; Cutler \& Lindblom 1987):
\be
\eta=1.1\times 10^{17}T_9^{-2}\left({\rho\over\rho_{\rm nuc}}\right)^{9/4}
~{\rm g/(cm~s)},
\ee
where $T_9$ is the temperature in units of $10^9$~K, and $\rho_{\rm nuc}
=2.8\times 10^{14}$~g~cm$^{-3}$.
The bulk viscosity results from the tendency to achieve chemical 
equilibrium via the modified URCA process during the perturbation
(Sawyer 1989):
\be
\zeta=4.8\times 10^{24}T_9^6\left({\rho\over\rho_{\rm nuc}}\right)^2
\left({\omega_r\over 1~{\rm s}^{-1}}\right)^{-2}~{\rm g/(cm~s)}.
\ee
We have not been able to derive an explicit expression for $(dE/dt)_{\rm visc}$
from the g-mode wavefunctions to facilitate convenient 
numerical computation of the viscous damping rate. We shall therefore be 
contented with order of magnitude estimates. The mode damping rate
due shear viscosity is approximately given by the rate at which momentum 
diffuses across a mode wavelength:
\be
{1\over \tau_{\rm shear}}\sim {\eta\over \rho L^2}
\simeq 10^{-2}L_1^{-2}T_9^{-2}\left({\rho\over\rho_{\rm nuc}}\right)^{5/4}
~{\rm yr}^{-1},
\label{shearrate}
\ee
where $L=10L_1$~km is the characteristic wavelength of the g-mode. 
Similarly, the mode damping rate due to bulk viscosity is estimated as
\be
{1\over\tau_{\rm bulk}}\sim 0.6\,L_1^{-2}
T_9^6\left({\rho\over\rho_{\rm nuc}}\right)
\left({\nu_r\over 150~{\rm Hz}}\right)^{-2}~{\rm yr}^{-1},
\label{bulkrate}\ee
(where $\nu_r$ is the mode frequency in the rotating frame).
The length scale of the mode is somewhat less than $10$~km, 
as comparison with the numerical results for nonrotating g-modes
(Lai 1994) indicates that eq.~(\ref{shearrate}) (with $L_1=1$)
underestimates $1/\tau_{\rm shear}$ by a factor of ten, while
eq.~(\ref{bulkrate}) underestimates $1/\tau_{\rm bulk}$ by a factor of 
three. When the star rotates faster than the 
mode frequency at zero rotation, the transverse wavelength of the mode
decreases as $\sim 2R/q$ (for $q\gg 1$), and the radial wavelength also
tends to be smaller (see Fig.~3). For the rotation rate of interest in 
this paper ($\Omega_s$ is at most a few times $\omega^{(0)}$), 
the mode wave function is only modestly changed by rotation, 
and we expect that rotation modifies the 
viscous damping rates by at most a factor of a few. 
Also note that
the mode frequency (in the rotating frame) $\nu_r$ increases
with increasing spin frequency (e.g., for model UU, we have $\nu_r=296$~Hz
when $\nu_s=300$~Hz), and this tends to reduce our estimate for
$\tau_{\rm bulk}^{-1}$. In any case, the uncertainty in our estimate
of viscous damping rate is probably comparable to the uncertainly in 
our understanding of the microscopic viscosity of neutron star matter. 

The total damping rate of the g-mode is given by
\be
{1\over \tau}={1\over\tau_{\rm gr}}+{1\over\tau_{\rm shear}}+
{1\over\tau_{\rm bulk}}.
\ee
The mode is unstable when $\tau^{-1}<0$. Despite the uncertainty in our
estimates for $\tau_{\rm shear}$ and $\tau_{\rm bulk}$, by comparing
Fig.~4 with eqs.~(\ref{shearrate}) and (\ref{bulkrate}) we can readily
conclude that for $T\go 10^9$~K the g-mode instability is suppressed
by the bulk viscosity, while for $T\lo 10^9$~K it is likely to be suppressed
by the shear viscosity. For a narrow range of neutron star temperatures, 
around a few times $10^8$~K to $10^9$~K (depending on the detail of
viscous dissipation), the g$_1$ mode may be unstable due to gravitational 
radiation. To achieve this instability, the rotation rate
of the star must be at least a factor of two to three greater than 
the mode frequency at zero rotation (see Fig.~4). 

So far we have only focused on the $j=m=2$ g$_1$ mode. 
As the order of the g-modes increases,
both the mode frequency and $\delta D_{lm}$ decreases, 
which tends to reduce $|\tau_{\rm gr}^{-1}|$. 
Also, the viscous mode damping rate increases with 
increasing mode order.  Therefore we expect that 
the CFS instabilities of higher-order g-modes are completely 
suppressed by viscous dissipations.

\section{Viscosity Driven Instability?}

For an incompressible rotating Maclaurin spheroid, the $l=-m=2$ f-mode
becomes unstable due to viscosity at the same point as the CFS instability
of the $l=m=2$ mode (Chandrasekhar 1987). The $l=-m>2$ viscous 
instability occurs at higher rotation rates. Recent studies
indicate the instability occurs only for neutron stars with very stiff
equation of state (Bonazzola et al.~1997).

\begin{figure}
\vskip -1.2 truecm
\epsfxsize=10.2cm
\epsfysize=10.2cm
\hbox{\hskip -1.0 truecm \epsfbox{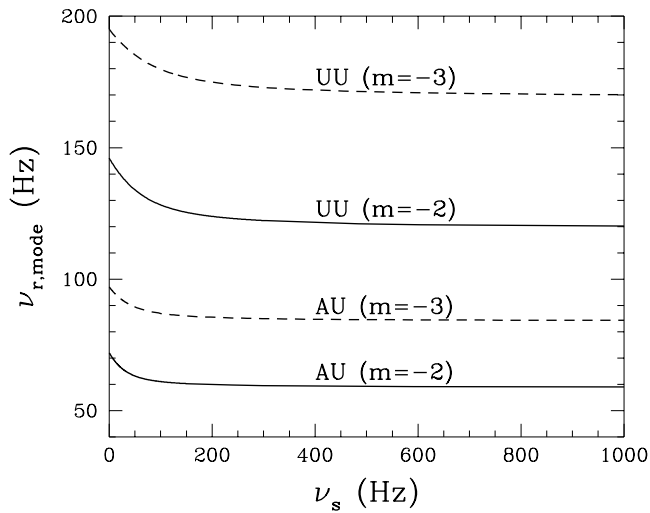}}
\vskip -1.9 truecm
\caption{The frequencies (in the rotating frame) of the $j=-m=2$ g$_1$
modes for the two neutron star models (UU and AU) as functions of
the spin frequency. For comparison, the dashed lines show the results for the
$j=-m=3$ g$_1$ modes.
}
\end{figure}

The viscous instability sets in when the mode frequency $\omega_r$ 
in the rotating frame goes through zero. Figure 5
shows the frequencies of the $j=-m=2$ modes for the two neutron star models
(UU and AU). For comparison, the frequencies of the $j=-m=3$ modes are also
shown. All these modes follow the approximate WKB scaling relation
$\omega_r\propto\sqrt{\lambda}$ (see Sect.~3).
We see that while $\omega_r$ is reduced by rotation, it never goes to
zero, up to the maximum possible rotation 
rate.\footnote{Note that as the rotation rate approaches the maximum value,
our calculation of the mode frequency breaks down since the
rotational distortion becomes significant.}
We therefore conclude that viscosity-driven instability is absent 
for g-modes.

\section{Discussion}

We have shown in this paper that g-modes in an inviscid neutron star are
susceptible to the CFS instability when the
rotation rate is comparable to or greater than the mode frequencies. 
However, viscous dissipation tends to suppress the instability
except possibly in a narrow range of temperatures around $10^9$~K. 

Comparing to f-modes, the g-mode instability has the advantage that
it sets in at a lower rotation rate. Comparing to r-modes, 
the g-mode instability is much weaker.
Calculations (Lindblom et al.~1998; Andersson et al.~1998a)
indicate that at spin frequency of $300$~Hz, 
the growth rate of r-mode is approximately $10^3$~yr$^{-1}$, more
than three orders of magnitude larger than $1/\tau_{\rm gr}$ for the
g-modes studied in this paper. Thus for isolated neutron stars,
the r-mode instability plays an much more important role than the g-mode
in determining the spin evolution of the stars (Owen et al.~1998;
Andersson et al.~1998b). 

However, one should not dismiss possible importance of neutron star 
g-modes and their stabilities in other situations. In 
a merging neutron star binary, g-modes can be resonantly excited
by the companion star, with the consequence of 
changing the orbital decay rate (Reisenegger \& Goldreich 1994; 
Lai 1994). Rotation may significantly change
the strength of the resonance (Lai 1997), giving rise to 
measurable effect in the gravitational wave form (Ho \& Lai 1999).
In another situation, when the neutron star is surrounded by
a disk, tidal coupling between the stellar g-modes
and disk density waves can drive the the unstable modes, 
which may affect the stellar rotation. By contrast, r-modes
possess no mass quadrupole moment, and the coupling with disk 
is mediated only through higher order multipole moments and thus
is expected to be weak.  We wish to study some of these issues
in a future paper.

\section*{Acknowledgments}

Part of this paper was completed while the author was visiting 
the Aspen Centre for Physics during May-June of 1998. 
He thanks the referee for constructive comments which improved
this paper, and acknowledges support from a NASA ATP grant, 
and a research fellowship from the Alfred P. Sloan Foundation.


\label{lastpage}

\begin{thebibliography}{99}

\bibitem{} Andersson, N., 1998, ApJ, 502, 708

\bibitem{} Andersson, N., Kokkotas, K.D., Schutz, B.F., 1998a, ApJ,
510, 846 

\bibitem{} Andersson, N., Kokkotas, K.D., Stergioulas, N., 1998b, 
preprint astro-ph/9806089.

\bibitem{} Baym, G., Bethe, H.A, \& Pethick, C.J. 1971, Nucl. Phys. A175,
225

\bibitem{} Bildsten, L. 1998, ApJ, 501, L89

\bibitem{} Bildsten, L., \& Cumming, A. 1998, ApJ, 506, 842

\bibitem{} Bildsten, L., Cutler, C., 1995, ApJ, 449, 800

\bibitem{} Bildsten, L., Ushomirsky, G., Cutler, C., 1996, ApJ, 460, 827

\bibitem{} Bonazzola, S., Frieben, J., Gourgoulhon, E., 1998, 
A\&A, 331, 280
%

\bibitem{} Chandrasekhar, S., 1970, Phys. Rev. Lett., 24, 611

\bibitem{} Chandrasekhar, S., 1987, Ellipsoidal Figures of Equilibrium
(Dover Pub.: New York)

\bibitem{} Chapman, S., Lindzen, R.S., 1970, Atmospheric Tides (Gordon and
Breach: New York)

\bibitem{} Cutler, C., Lindblom, L., 1987, ApJ, 314, 234

\bibitem{} Cutler, C., Lindblom, L., 1992, ApJ, 385, 630

\bibitem{} Finn, L.S., 1987, MNRAS, 227, 265

\bibitem{} Flowers, E., Itoh, N., 1979, ApJ, 230, 847.

\bibitem{} Friedman, J.L., Morsink, S.M., 1998, ApJ, 502, 714

\bibitem{} Friedman, J.L., Schutz, B.F., 1978a, ApJ, 221, 937

\bibitem{} Friedman, J.L., Schutz, B.F., 1978b, ApJ, 222, 281

\bibitem{} Ho, W.C.G., Lai, D. 1999, MNRAS, in press (astro-ph/9812116)

\bibitem{} Ipser, J.R., Lindblom, L., 1990, ApJ, 355, 226

\bibitem{} Ipser, J.R., Lindblom, L., 1991, ApJ, 373, 213

\bibitem{} Lai, D., 1994, MNRAS, 270, 611

\bibitem{} Lai, D., 1997, ApJ, 490, 847

\bibitem{} Levin, Y. 1998, astro-ph/9810471

\bibitem{} Lindblom, L., 1995, ApJ, 438, 265

\bibitem{} Lindblom, L., Owen, B.J., Morsink, S.M., 1998, Phys. Rev. Lett.,
80, 4843 

\bibitem{} Longuet-Higgins, M.S., 1967, Phil. Trans. R. Soc. London A, 262, 511

\bibitem{} McDermott, P.N., \& Taam, R.~E. 1987, ApJ, 318, 278

\bibitem{} McDermott, P.N., Van Horn, H.M., Hansen, C.J., 1988, ApJ, 
325, 725

\bibitem{} Owen, B.J., Lindblom, L., Cutler, C., Schutz, B.F., Vecchio, A.,
Andersson, N., 1998, Phys. Rev. D 5808, 4020

\bibitem{} Papaloizou, J., Pringle, J.E. 1978, MNRAS, 182, 423

\bibitem{} Provost, J., Berthomieu, G., Rocca, A., 1981, A\&A, 94, 126

\bibitem{}
Reisenegger, A, 1994, ApJ, 432, 296

\bibitem{}
Reisenegger, A, Goldreich, P., 1992, ApJ, 395, 240

\bibitem{}
Reisenegger, A, Goldreich, P., 1994, ApJ, 426, 688

\bibitem{} Saio, H., 1982, ApJ, 256, 717

\bibitem{} Stergioulas, N., \& Friedman, J.L., 1998, ApJ, 492, 301

\bibitem{} Strohmayer, T.~E., 1993, ApJ, 417, 273.

\bibitem{} Strohmayer, T.~E., \& Lee, U. 1996, ApJ, 467, 773

\bibitem{} Thorne, K.S., 1980, Rev. Mod. Phys., 52, 299

\bibitem{} Unno, W. et al., 1989, Nonradial Oscillation of Stars
(Univ. of Tokyo Press)

\bibitem{} Wiringa, R.B., Fiks, V., Fabrocini, A., 1988, Phys. Rev. C, 38, 1010

\bibitem{} Yoshida, S., Eriguchi, Y., 1995, ApJ, 438, 830


\end{thebibliography}
\end{document}